\documentclass[sigconf]{sig-alternate-per}

\usepackage[utf8]{inputenc}

\title{Learning to Dynamically Select Cost Optimal Schedulers in Cloud Computing Environments}
\author{Shreshth Tuli$^1$, Giuliano Casale$^1$, Nicholas R. Jennings$^{1, 2}$ \\ $^1$Imperial College London, UK , $^2$Loughborough University, UK}

\usepackage[utf8]{inputenc}
\usepackage{amsmath,amssymb,amsfonts}
\usepackage{mathtools}
\usepackage[style=numeric]{biblatex}
\usepackage[noindentafter]{titlesec}
\usepackage{graphicx}
\usepackage{color}
\usepackage{hyperref}
\usepackage{subcaption}
\usepackage{booktabs}
\graphicspath{ {images/} }
\usepackage{dsfont}

\DeclarePairedDelimiter\norm{\lVert}{\rVert}
\DeclareMathOperator*{\argmin}{arg\,min}

\begin{document}

\maketitle

\begin{abstract}
The operational cost of a cloud computing platform is one of the most significant Quality of Service (QoS) criteria for schedulers, crucial to keep up with the growing computational demands. Several data-driven deep neural network (DNN)-based schedulers have been proposed in recent years that outperform alternative approaches by providing scalable and effective resource management for dynamic workloads. However, state-of-the-art schedulers rely on advanced DNNs with high computational requirements, implying high scheduling costs. In non-stationary contexts, the most sophisticated schedulers may not always be required, and it may be sufficient to rely on low-cost schedulers to temporarily save operational costs. In this work, we propose MetaNet, a surrogate model that predicts the operational costs and scheduling overheads of a large number of DNN-based schedulers and chooses one on-the-fly to jointly optimize job scheduling and execution costs. This facilitates improvements in execution costs, energy usage and service level agreement violations of up to 11\%, 43\% and 13\% compared to the state-of-the-art methods.
\end{abstract}

\section{INTRODUCTION}
\label{sec:introduction}
The onset of the Artificial Intelligence (AI) and Deep Learning (DL) era has led to a recent shift in computation from hand-encoded algorithms to data-driven solutions for resource management in cloud systems~\cite{cai2016iot}. This has given rise to efficiently harnessing the data processing capacities of multiple devices and provides services at scale with high Quality of Service (QoS). However, the increasing computational demands of modern Internet of Things (IoT) applications makes it crucial to curtail the operational costs of cloud machines. This calls for efficient resource management schemes, such as task scheduling policies, to execute workloads on cloud systems within tight cost budgets. AI and DL offer promising solutions that rely on accurate surrogate models that are inexpensive to evaluate at run-time.

\textbf{Background and Motivation.} In recent years, state-of-the-art resource management solutions, which particularly focus on optimal placement of tasks on cloud virtual machines (VMs), have increasingly explored data-driven DL methods~\cite{graf, semidirect, tuli2021gosh, tuli2021cosco, decisionnn}. Such methods typically rely on a trace of resource utilization characteristics and QoS metrics of tasks and cloud hosts and are referred to as \textit{trace-driven} schedulers. They utilize such traces to train a deep neural network (DNN) to estimate a set of Quality of Service (QoS) parameters and run optimization strategies to find the best scheduling decision for each incoming task. However, most prior work run inference using such trace-driven DNN models on a broker node to generate scheduling decisions, while the incoming tasks are executed on worker nodes~\cite{tuli2021cosco}. In such cases, having a dedicated broker is often cost inefficient due to the idle times between decisions in discrete-time control settings, wherein the task placement decisions are taken at fixed scheduling intervals~\cite{tuli2021cosco}. To tackle this, we resort to paradigms such as Function as a Service (FaaS) that allow execution of DL schedulers as serverless functions, only incurring costs for the inference run time of each DNN model.

\textbf{Contributions.} In this work, to leverage the recent advantages brought by DL in task scheduling and build a policy agnostic solution, we choose a set of state-of-the-art schedulers. This work aims to solve the meta-problem of on-the-fly selection of scheduling policies for a cloud computing environment wherein the incoming tasks are executed on worker nodes and scheduling decisions are run as serverless functions. To solve this meta-problem, we develop the proposed solution that we call MetaNet. It uses a DNN as a surrogate model to predict the task execution costs and scheduling time for each policy. We then select the most cost-efficient policy at each scheduling interval using the online estimates generated by this surrogate. As we dynamically update the policy to tradeoff between simple and sophisticated DL schedulers, we reduce overall operational costs, energy consumption and Service Level Agreement (SLA) deadline violation rates by up to 11\%, 43\% and 13\% respectively compared to state-of-the-art schedulers.  

\section{BACKGROUND AND RELATED WORK}

Recent work in scheduling for cloud computing environments has demonstrated that AI-based solutions are not only faster, but can also scale efficiently compared to traditional heuristic and optimization techniques~\cite{semidirect, tuli2021gosh, tuli2021cosco, decisionnn}.

\textbf{Evolutionary Optimization.} This class of methods forecasts QoS using non-DL solutions such as ARIMA~\cite{arima} or DL based models such as Long-Short-Term-Memory (LSTM) neural networks~\cite{lstm}. It then applies evolutionary search strategies such as Ant Colony Optimization (ACO)~\cite{aco}, to converge to a locally optimal scheduling decision.

\textbf{Surrogate Optimization.} This class of methods uses differentiable function approximators, particularly neural networks, to act as surrogates of the QoS for a future state of the system. For instance, GRAF~\cite{graf} uses a graph neural network (GNN) as a surrogate model to predict service latencies and operational costs for a given scheduling decision and uses gradient-based optimization to minimize the service latencies or execution costs. To do this, it uses the concept of neural network inversion~\cite{neuralinversion}, wherein the method evaluates gradients of the objective function with respect to inputs and runs optimization in the input space. Other methods, such as Decision-NN~\cite{decisionnn}, GOBI~\cite{tuli2021cosco} and its second-order generalization GOSH~\cite{tuli2021gosh}, combine the prediction and optimization steps by modifying the loss function to train and optimize in tandem.

\section{PROPOSED METHOD}

\begin{figure}
    \centering \setlength{\belowcaptionskip}{-15pt}
    \includegraphics[width=\linewidth]{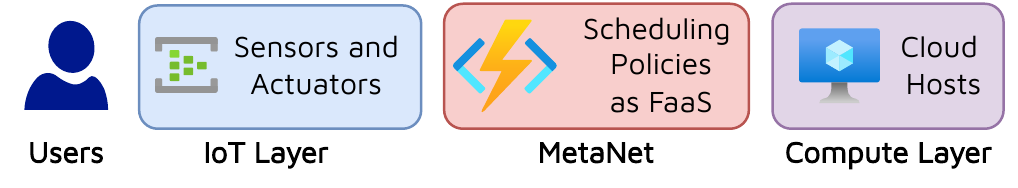}\vspace{-6pt}
    \caption{MetaNet System Model.}
    \label{fig:system}
\end{figure}

\textbf{System Model.} In this work, we target a standard heterogeneous cloud computing environment where all nodes are in the same Wide Area Network (see Figure~\ref{fig:system} for an overview). Tasks are container instances that need to be processed, generated from an IoT layer and transferred to the compute nodes via gateway devices. We do not have any cloud broker; instead, we run the scheduling policies as FaaS functions on a serverless platform such as AWS Lambda or Azure Serverless. As is common in prior work~\cite{aco, semidirect, decisionnn}, we consider a discrete-time control problem, \textit{i.e.}, we divide the timeline into fixed-size execution intervals (of $\Delta$ seconds). We denote the overall cost of the system amortized by the number of completed tasks in the $t$-th interval as $\phi_t$. This includes the operational costs of the worker nodes as well as that of running the schedulers as serverless functions.

\textbf{MetaNet Methodology.} We create a surrogate model $f_\theta$ that is a DNN with parameters $\theta$.  We consider a set of scheduling policies $\mathcal{P}$ of size $q$. Given a system state at the start of the $t$-th interval, each scheduler produces a scheduling decision for this interval. The state is denoted by $C_{t-1}$ and includes the resource (CPU, RAM and disk) utilization characteristics of the active workloads in the system $W_{t-1}$, resource utilization of cloud hosts $H_{t-1}$ and scheduling decision of the previous interval $S_{t-1}$. Thus,  $C_{t-1} = [W_{t-1}, H_{t-1}, S_{t-1}]$. Now, $f_\theta$ estimates the average execution cost (sum of the cost of task execution and serverless run) for each scheduler $p^k \in \mathcal{P}$ in the $t$-the interval as $\hat{\phi}^k_t$.  The collection of $\hat{\phi}^k_t$ for all $k$ in $I_t$ is denoted by $\hat{\phi}_t$. In such a case, our problem can be formulated as
\begin{equation}
\label{eq:problem}
\begin{aligned}
& \underset{\theta}{\text{minimize}} 
& & \sum_{t=1}^T \phi^\pi_t\\
& \text{subject to}
& & S_t = p^k(t), \forall\ t \\
&&& \pi = \argmin_k \hat{\phi}^k_t,  \forall\ t  \\
&&& \hat{\phi}_t = f_\theta(W_{t-1}, H_{t-1}, S_{t-1}), \forall\ t,
\end{aligned}
\end{equation}
where $f_\theta$ is a surrogate of the average execution cost.

\begin{figure}[t]
    \centering \setlength{\belowcaptionskip}{-15pt}
    \includegraphics[width=\linewidth]{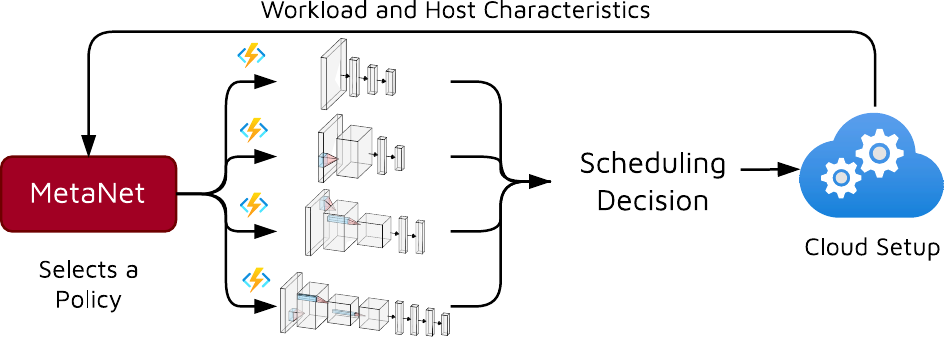}
    \caption{MetaNet Pipeline.}
    \label{fig:overall}
\end{figure}

\textbf{Surrogate Model and Training.} We use a feed-forward neural model that takes as input the state of the cloud system $C_{t-1} = [W_{t-1}, H_{t-1}, S_{t-1}]$ and outputs a single scalar. We use a fully connected neural network with 3 layers, each with 64 nodes and $\mathrm{ReLU}$ activation, except the last layer where we use the $\mathrm{Sigmoid}$ activation to bring the output in the range $(0, 1)$. Thus, for any input $(W_{t-1}, H_{t-1}, S_{t-1})$, the complete neural network can be described as
\begin{equation}
    \hat{\phi} = f_\theta(W_{t-1}, H_{t-1}, S_{t-1}).
\end{equation}

To train the above described MetaNet neural network $f_\theta$, we collect traces from a cloud computing environment. To collect data for training, we execute the scheduling policies $\mathcal{P}$, each for $\Gamma$ scheduling intervals, and collect a dataset as
\begin{equation}
    \label{eq:dataset}
    \Lambda = \{(k, W_{t-1}, H_{t-1}, S_{t-1}, \phi^k_t, \omega^k_t)\}_{t = 1}^{\Gamma \cdot q},
\end{equation}
where $q$ is the number of scheduling policies in $\mathcal{P}$. We initialize $W_0$ as $H_0$ zero-matrices and $S_0$ as an empty graph. We find the maximum cost from the dataset as $\phi^k_{max}$ for each scheduler $p^k$. This allows us to denormalize the neural network output and bring it to the same range as the one in the dataset. We then train the model using the loss function
\begin{equation}
    L = \norm{\phi^k - \phi^k_{max} \cdot \hat{\phi}^k}_{2}.
\end{equation}

\textbf{Dynamic Policy Selection Using MetaNet.} We first store the saved surrogate model $f_\theta$ into a central network-attached-storage (NAS) that is accessible to all worker nodes. At start the scheduling interval $I_t$, we get the workload and host characteristics $W_{t-1}, H_{t-1}$ with the scheduling decision $S_{t-1}$.  As we do not have any broker node in our setup, we run the MetaNet model in the worker node with the least CPU utilization. On this worker node we use the surrogate model to predict cost and decide the scheduling policy as
\begin{equation}
    \label{eq:decide}
    p^\pi, \text{ s.t. } \pi = \argmin_k \hat{\phi}^k_t.
\end{equation}
Overall, MetaNet selects a scheduling policy on-the-fly as per the system states. To do this, at the start of each scheduling interval, it predicts the task execution cost and scheduling time of each policy. It then selects the one with the minimum cost estimate. This policy optimizes the scheduling decision, initialized as $S_{t-1}$, to generate $S_t$. The complete pipeline is summarized in Figure~\ref{fig:overall}. We also tune $f_\theta$ with each new datapoint to adapt to non-stationary settings.

\section{PERFORMANCE EVALUATION}

\textbf{Setup.} We consider the complete set of hosts $\mathcal{H}$ to be static with time as is common in prior work for a fixed cloud platform~\cite{semidirect}. We use different VM types from the Microsoft Azure Platform, \textit{i.e.}, 60 \texttt{Azure B2s} with a dual-core CPU and 4GB RAM (in UK-South), 20 \texttt{Azure B4ms} with a quad-core CPU and 16GB RAM and 20 \texttt{Azure B8ms} with an octa-core CPU and 32 GB RAM (in East-US). To save on costs, we define a host to be active when the CPU utilization is $> 0\%$. We utilize the \texttt{Azure Automation} (\url{https://azure.microsoft.com/services/automation/\#overview}) service to hibernate and resume VMs based on their CPU utilization.

\begin{figure}[t]
    \centering \setlength{\belowcaptionskip}{-13pt}
    \includegraphics[width=\linewidth]{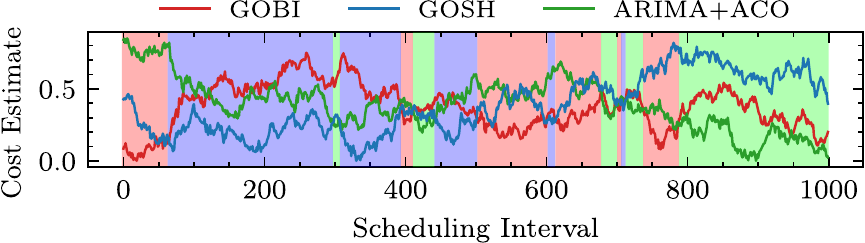} 
    \caption{Predicted costs (line plots) and dynamic scheduler selection (background color) in MetaNet.}
    \label{fig:vis}
\end{figure}

\textbf{Workloads.} To generate the tasks in our system, we use the \textit{AIoTBench} applications~\cite{luo2018aiot}. AIoTBench is an AI based cloud computing benchmark suite that consists of various real-world computer vision application instances. The seven specific application types correspond to the neural networks they utilize. These include three typical heavy-weight networks: ResNet18, ResNet34, ResNext32x4d, as well as four light-weight networks: SqueezeNet, GoogleNet, MobileNetV2, MnasNet. At the start of each scheduling interval, we create new tasks, sampled uniformly from the seven applications, where the number of new tasks comes from a Poisson distribution with rate $\lambda = 1.2$.

\textbf{Baselines.} We compare MetaNet against seven baselines, which also form our policy set $\mathcal{P}$. We integrate the ACO algorithm with two demand forecasting methods: AutoARIMA and LSTM, and call these ARIMA+ACO and LSTM+ACO. We also include Decision-NN, Semi-Direct and GRAF, GOBI and GOSH. This makes the size of the policy set $q = 7$. We also use a Multi-Armed Bandit (MAB) model using Upper-Confidence-Bound exploration~\cite{kuleshov2014algorithms} and a Deep-Q-Network (DQN)~\cite{li2017deep} that choose a policy based on pre-trained models using data $\Lambda$.

\textbf{Implementation and Training.} We use implement MetaNet on the COSCO framework~\cite{tuli2021cosco}.  We collect the dataset $\Lambda$ by executing all baselines for $\Gamma = 100$ intervals. Similarly, we execute all approaches for $T = 1000$ scheduling intervals to generate QoS scores, with each interval being $\Delta = $ 10 seconds long, giving a total experiment time of nearly 2 hours 46 minutes for each method. 

\textbf{Visualization of MetaNet.} Figure~\ref{fig:vis} visualizes the MetaNet approach running on the setup and workloads described above. The x-axis denotes the scheduling interval and the y-axis denotes the cost estimate $\hat{\phi}^k_t$ for the three most frequently selected policies in $\mathcal{P}$ for readability: ARIMA+ACO, GOBI and GOSH. The highlighted bands indicate the selected scheduling policy. The selected policy corresponds to the one with the least estimated cost. The cost estimates are non-stationary, further corroborating the need for dynamic selection of the scheduling policies in volatile workload and host setups. 

\textbf{Comparison with Baselines.} Table~\ref{tab:comparison} shows the results of our experiments (cost in USD, energy in KW-hr, resp. time in seconds). Across all metrics, MetaNet outperforms the baselines. MetaNet improves the energy consumption by allocating tasks, \textit{i.e.}, to the same hosts to minimize execution costs and consequently the active hosts in the system. This is shown by the highest CPU utilization of MetaNet, \textit{i.e.}, 87.4\%. MetaNet also gives the lowest response time and consequently the lowest SLA violation rates. Dynamic optimization baselines perform poorly due to the stateless assumption in MAB that ignores environment dynamism, and DQN being slow to adapt in volatile settings~\cite{tuli2021cosco}.

\begin{table}[t]
    \centering \setlength{\belowcaptionskip}{-12pt}
    \caption{Comparison with the baseline.\vspace{-4pt}}
    \resizebox{\linewidth}{!}{
    \begin{tabular}{@{}lccccc@{}}
    \toprule 
    \textbf{Model} & \textbf{Cost} & \textbf{Energy} & \textbf{Resp. T.} & \textbf{SLA V.} & \textbf{CPU \%}\tabularnewline
    \midrule
    ARIMA+ACO & 0.794 & 5.432 & 20.391 & 0.241 & 62.4\tabularnewline
    LSTM+ACO & 0.862 & 5.217 & 16.921 & 0.212 & 76.5\tabularnewline
    Decision-NN & 1.072 & 4.224 & 12.255 & 0.211 & 80.2\tabularnewline
    Semi-Direct & 1.021 & 4.095 & 17.092 & 0.131 & 72.2\tabularnewline
    GRAF & 0.993 & 4.228 & 14.722 & 0.127 & 76.1\tabularnewline
    GOBI & 0.752 & 3.827 & 14.293 & 0.122 & 80.2\tabularnewline
    GOSH & 0.911 & 3.267 & 13.292 & 0.118 & 84.3\tabularnewline
    \midrule
    MAB & 0.874 & 2.921 & 13.921 & 0.121 & 79.9\tabularnewline
    DQN & 0.907 & 3.121 & 13.877 & 0.119 & 80.4\tabularnewline
    MetaNet & \textbf{0.667} & \textbf{2.029} & \textbf{11.273} & \textbf{0.102} & \textbf{87.4}\tabularnewline
    \bottomrule
    \end{tabular}
    } \vspace{-6pt}
    \label{tab:comparison}
\end{table}

\section{Conclusions}
This work presents MetaNet, which dynamically selects scheduling policies for cost-efficient task processing in cloud systems. Future work would explore other DNNs as surrogates and extend MetaNet to other types of resource management decisions such as dynamic VM provisioning and autoscaling.


\AtNextBibliography{\small}
\printbibliography

\end{document}